\documentclass[runningheads]{llncs}
\usepackage{graphicx}
\usepackage{caption}
\usepackage{subcaption}
\usepackage{times}

\begin{document}
\title{xPACE and TASC Modeler: \\Tool support for data-driven context modeling
\thanks{This work was partially supported by CNPq, Brazil.}
}
%
%
\author{Rodrigo Falc\~{a}o\inst{1}\orcidID{0000-0003-1222-0046} \and
Rafael King\inst{1}\orcidID{0000-0001-7107-5390} \and
Ant\^{o}nio L\'{a}zaro Carvalho\inst{2}\orcidID{0000-0002-0013-4565}}
\authorrunning{R. Falc\~{a}o et al.}
%
\institute{Fraunhofer IESE, Kaiserslautern (Germany)
\and
Computer Science Graduate Program, Federal University of Bahia, Salvador (Brazil)
}
\maketitle              
\begin{abstract}
From a requirements engineering point of view, the elicitation of context-aware functionalities calls for context modeling, an early step that aims at understanding the application contexts and how it may influence user tasks. In practice, however, context modeling activities have been overlooked by practitioners due to their high complexity. To improve this situation, we implemented xPACE and TASC Modeler, which are tools that support the automation of context modeling based on existing contextual data. In this demonstration paper, we present our implementation of a data-driven context modeling approach, which is composed of a contextual data processor (xPACE) and a context model generator (TASC Modeler). We successfully evaluated the results provided by the tools in a software development project.
\end{abstract}
\section{Introduction}

Computers have become increasingly ubiquitous, and we are witnessing the rise of applications, sensors, and networks that together deliver smart behaviors to users. Context awareness plays a key role in this game as a core characteristic of ubiquitous computing \cite{poslad2011ubiquitous} \cite{spinola2012towards} and is frequently behind the perceived ``intelligence'' of modern software solutions\cite{pinheiro2018supporting}.

Context-aware functionalities are functionalities that consider context to produce a certain system behavior, typically an adaptation or recommendation. From a requirements engineering point of view, the elicitation of context-aware functionalities requires context modeling, which is an early step that involves identification of contextual elements, analysis of accessibility (i.e., which contextual elements have available sources from which their values can be read), analysis of the relevance of these contextual elements for user tasks of interests, and analysis of combinations of contextual elements for these user tasks \cite{falcao2017improving}. The analysis of relevance and combinations can be challenging, though: In a scenario with dozens of contextual elements, how to figure out which contextual elements, particularly in combination with each other, may influence a given user task? As the number of contextual elements increases, the number of potential combinations grows exponentially. In practice, these context modeling activities have been overlooked due to their high complexity: Practitioners regard them as time-consuming, non-intuitive, and error-prone \cite{falcao2021practical}. As a consequence, opportunities for discovering unexpected context-aware functionalities are missed.

We approached this problem by automating context modeling for the elicitation of context-aware functionalities. For this purpose, we designed a data-driven context modeling process (first introduced in \cite{falcao2017improving}) that has three parts. First, the user task of interest is analyzed; then, contextual data that relates to the user task is collected; and finally, at the core of the data-driven context modeling process, the contextual data is processed to identify relevant combinations of contextual elements -- and finally a context model is generated.

To come alive, the proposed automation in the process requires the implementation of two software components. One is the \textit{Contextual Data Processor}, which is responsible for analyzing a contextual dataset. Its output is used by the other component, the \textit{Context Model Generator}, which is responsible for creating the concrete context model that is expected to support requirements engineers in the elicitation of context-aware functionalities. In this paper, we present our implementations of these software components: \textit{xPACE} and \textit{TASC Modeler} \footnote{Source code available: https://github.com/rmfalcao/tasc-modeler-xpace\label{fn:source-code}}, and briefly discuss its application in a software development project at Fraunhofer IESE.

\section{Solution overview}

\begin{figure}[t]
\centerline{\includegraphics[width=\textwidth]{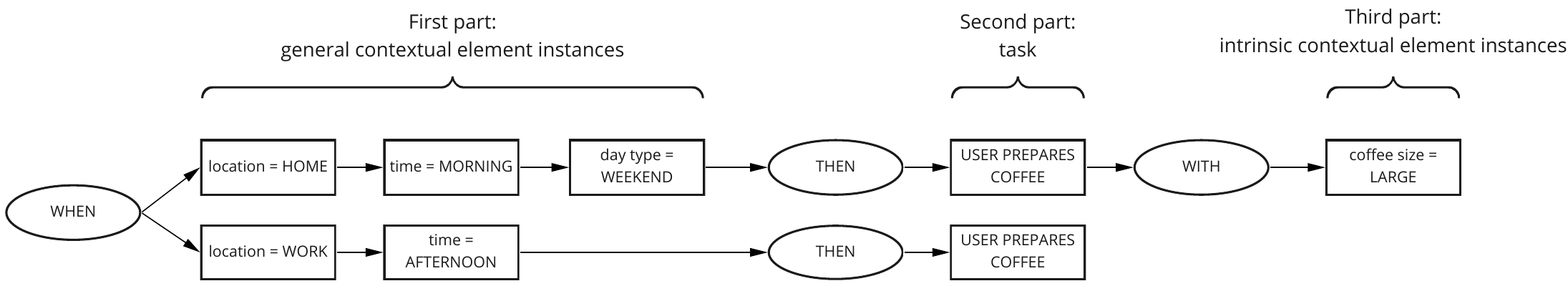}}
\caption{Example of a task-specific context model using TASC4RE.} \label{fig:coffee-example}
\end{figure}

Once a user task of interest has been chosen and corresponding contextual data has been collected, a contextual dataset is available. Then the requirements engineer uses TASC Modeler to generate the desired context model, providing the contextual dataset as input, in addition to a metadata file that configures the behavior of both the context model generator -- how the contexts will be presented in the model -- and the contextual data processor -- how the dataset should be analyzed. The detailed specification of the input files can be found with the source code. TASC Modeler forwards the dataset and the metadata to xPACE, which analyzes the data and returns the findings in a data structure named \textit{standardized task-specific contexts}, which is created to decouple the data analysis from the context model generation. Then TASC Modeler translates the standardized task-specific contexts provided by xPACE into a context model representation. In our case, the context model is represented as a directed acyclic graph with one root node. Each path from the root node towards a leaf describes how a context influences a user task of interest. Figure~\ref{fig:coffee-example} shows an example. Consider the user task ``Prepare a coffee''. Each of the two paths contains a set of instantiated contextual elements that, together, were found to influence the task (e.g., ``When location = WORK and time = AFTERNOON then user prepares coffee'', i.e., the context ``location =WORK and time = AFTERNOON'' influences the user task ``Prepare a coffee'', according to the model).

\begin{figure}
    \begin{subfigure}[b]{0.5\textwidth}
    \centering
    \includegraphics[width=\textwidth]{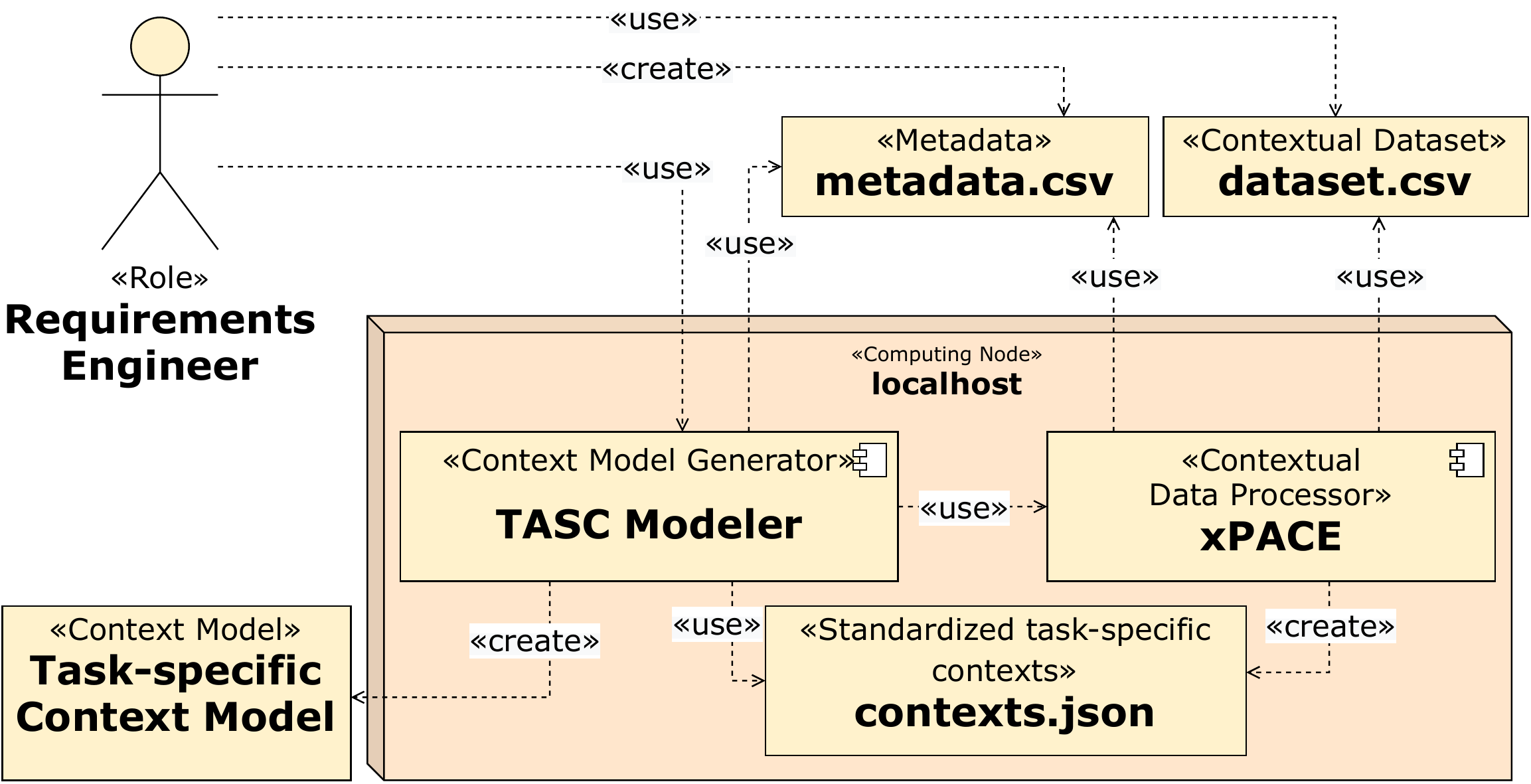}
    \caption{Component diagram of the solution.}
    \label{fig:component-diagram}
    \end{subfigure}
    \hfill
    \begin{subfigure}[b]{0.5\textwidth}
        \centering
        \includegraphics[width=\textwidth]{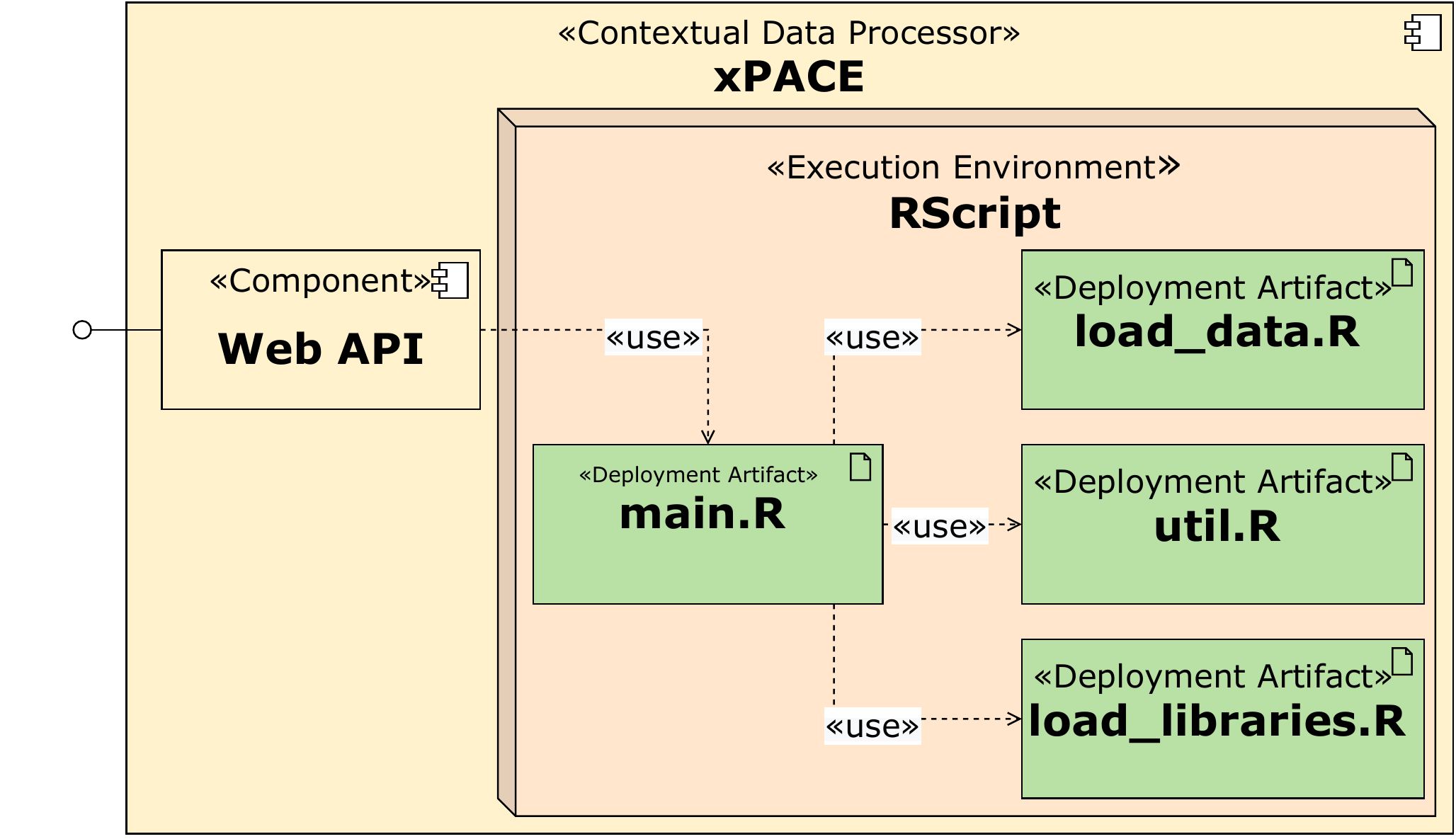}
        \caption{Detailed view of xPACE.}
        \label{fig:contextual-data-processor}
    \end{subfigure}
    \caption{The architecture of the solutions (functional view at runtime).}
    \label{fig:component-diagram-2}
\end{figure}

We implemented the dataset and the metadata artifacts as comma-separated value (CSV) files, whereas we used JSON for the standardized task-specific contexts file. Figure~\ref{fig:component-diagram} shows the component diagram of the solution.

TASC Modeler and xPACE are containerized and can be deployed and executed in any machine running the Docker Engine\footnote{https://docker.com}, an open-source containerization technology. The Docker container for starting TASC Modeler requires approx. 600 MB of RAM, whereas for xPACE, approx. 300 MB are required -- more memory may be needed depending on the dataset size.

\section{xPACE -- eXtended Pairwise Analysis of Contextual Elements}\label{implementation:sec:data-processor}

\begin{figure}[t]
    \centering
    \centerline{\includegraphics[width=\linewidth]{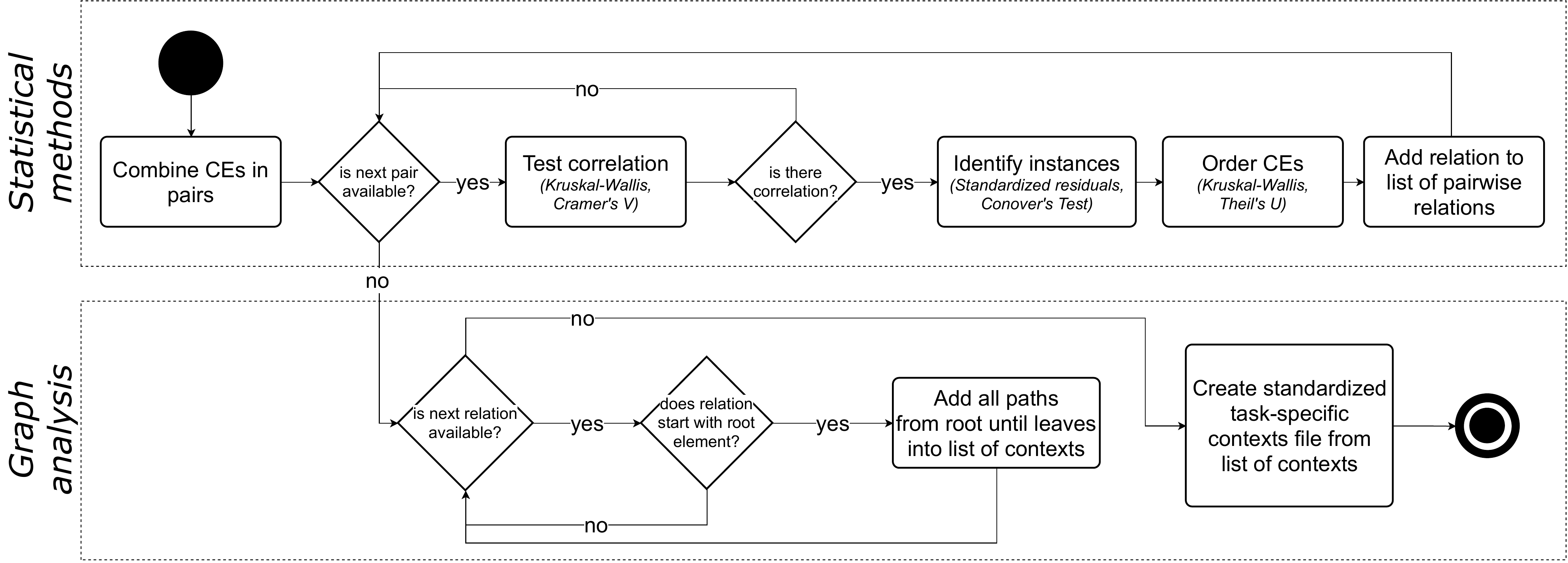}}
    \caption{Activity diagram illustrating the algorithm of xPACE.}
    \label{fig:main-algorithm}
\end{figure}

The \textit{eXtended Pairwise Analysis of Contextual Elements} (xPACE) was implemented using a strategy that can be divided into two parts. First, it uses statistical methods to search for correlations between pairs of contextual elements (CEs) in the contextual dataset. Whenever correlations are found, the algorithm identifies which contextual element instances (i.e., concrete values of the contextual elements) are correlated. This is necessary because it is not enough to know that two contextual elements relate to each other (e.g. ``location'' and ``time''), but it also needs to be know which instances relate to each other (e.g., ``at home'' and ``evening''). After that, the pair is ordered to express the direction of the relationship. After analyzing all pairs of contextual elements, we end up with a \textit{list of pairwise relations}, which is a set of directed pairwise relations among contextual element instances. The second part of the strategy takes the list of pairwise relations and builds a graph $G$ by treating each pair as an edge of the graph. When all pairs have been added to $G$, each path in the graph starting from a root node and ending on a leaf node will represent a relevant combination of contextual elements that were found by xPACE to influence the user task. Figure~\ref{fig:main-algorithm} contains an activity diagram that illustrates the data processor algorithm. The employed statistical methods are named in the corresponding steps.

The algorithm is implemented using the R language\footnote{https://www.r-project.org/}, and a Web API component implemented in Java provides handy access to it. The core takes as input the name of the user task and two files: a contextual dataset and a metadata file that describes the dataset; as output, it produces a standardized task-specific contexts file, which can be used by any context model generator that complies with its layout. Figure~\ref{fig:contextual-data-processor} shows the internal structure of xPACE. The principal file in the core is \texttt{main.R}. It contains the algorithm that we implemented to analyze contextual data to identify potential relevant contexts for the user task in focus. The other R files are helpers (``load\_libraries.R'' loads the necessary packages into the memory, ``util.R'' defines some functions, and ``load\_data.R'' reads the input files into the memory). 

\section{TASC Modeler}

The \textit{Task-specific Context Modeler} (TASC Modeler) is responsible for creating the graphical representation of the context model. The requirements engineer interacts directly with TASC Modeler, providing the contextual dataset, the metadata, and the task name as input. TASC Modeler then interfaces with a contextual data processor to provide the input data and receive the standardized task-specific context files as response. In our case, TASC Modeler uses xPACE, but it could be any other contextual data processor implementation able to generate the standardized task-specific context file via a REST API. TASC Modeler reads from a configuration file the information about which contextual data processor it should use.

We implemented TASC Modeler as a single-page application written in Typescript using the React Library\footnote{https://reactjs.org/}. When the application is loaded, the user is presented a form where they can provide the dataset file, the metadata file, and the name of the user task in focus. When the button ``Generate'' is pressed, the application sends the input data to xPACE through its Web API. Then xPACE returns its implementation of the standardized task-specific contexts file, which is used by TASC Modeler to build the context model. 

\begin{figure*}[]
    \centering
    \centerline{\includegraphics[width=\linewidth]{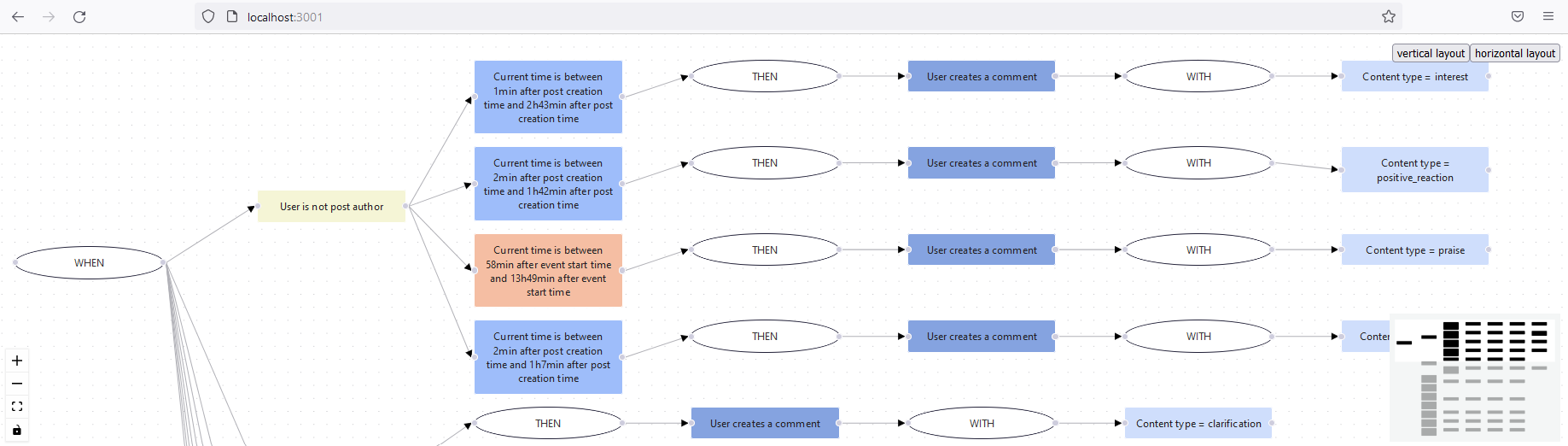}}
    \caption{Screenshot of TASC Modeler showing an excerpt of the task-specific context model.}
    \label{fig:cmg-diagram}
\end{figure*}

\section{Application and evaluation}

We used TASC Modeler and xPACE to support the elicitation of context-aware functionalities for DorfFunk\footnote{https://www.digitale-doerfer.de/unsere-loesungen/dorffunk/}, a live communication app with approx. 25,000 active users developed and maintained by Fraunhofer IESE. We chose a user task (``create a comment''), collected contextual data (approx. 56,000 tuples from 15 contextual elements), and used the tools. Figure~\ref{fig:cmg-diagram} shows a screenshot of TASC Modeler with an excerpt of the data-driven context model generated using DorfFunk data. The generated context model was evaluated in a controlled experiment with professional software engineers, where it showed its potential to support the identification of relevant contexts for given user tasks. All participants were asked to elaborate context-aware functionalities to improve the targeted user task. Participants of the treatment group received the data-driven context model, whereas participants of the control group received the list of contextual elements that were available to describe context-aware functionalities. On average, participants of the treatment group were able to elaborate context-aware functionalities that combine more contextual elements. They also stated that they found the context model valuable for supporting the elicitation of context-aware functionalities.

\begin{figure}[t]
    \begin{subfigure}[b]{0.5\textwidth}
    \centering
    \includegraphics[width=\textwidth]{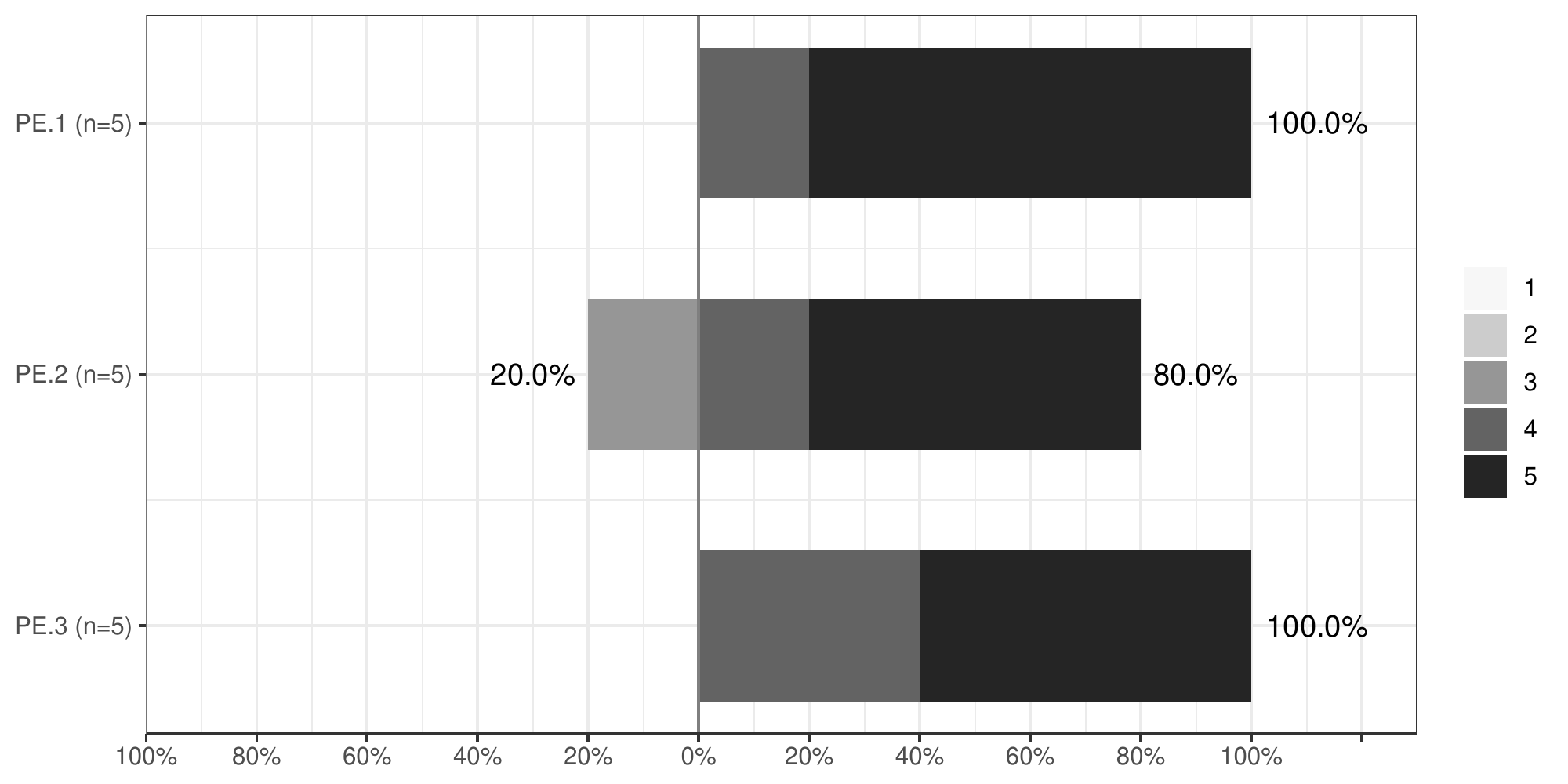}
    \caption{Performance expectation.}
    \label{fig:utaut-pe}
    \end{subfigure}
    \hfill
    \begin{subfigure}[b]{0.5\textwidth}
        \centering
        \includegraphics[width=\textwidth]{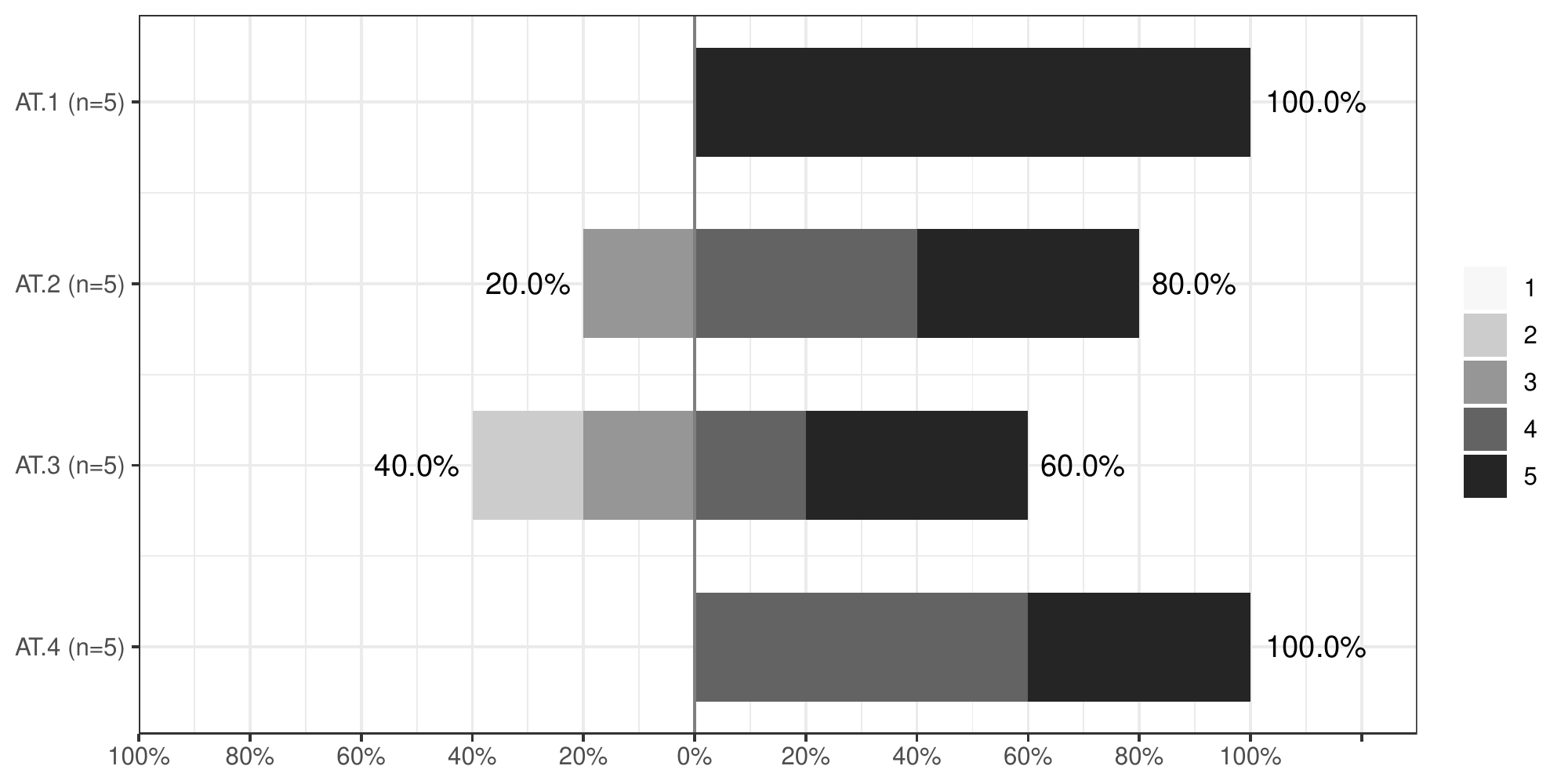}
        \caption{Attitude towards using technology.}
        \label{fig:utaut-at}
    \end{subfigure}
    \caption{Participants' assessment of the model's usefulness.}
    \label{fig:utaut}
\end{figure}

While the details of the experiment can be found in our previous paper \cite{falcao2021experiment}, we here reproduce some parts that concern the usage of the context model. Our hypothesis was: ``The data-driven context model is perceived by individuals as a useful instrument to support the elicitation of context-aware functionalities''. In order to test this hypothesis, we employed the UTAUT (Unified Theory of Acceptance and Use of Technology \cite{venkatesh2003user}). In Figure~\ref{fig:utaut}, the positive trend towards the acceptance of the context model can be noted.

\section{Conclusion}

Context modeling to support the elicitation of context-aware functionalities has been disregarded by practitioners due to its high complexity. To improve this scenario, we designed and implemented a data-driven context modeling process that automates the analysis of combinations of contextual elements that influence user tasks and the generation of the context model. The automation of this process is supported by two tools: the Contextual Data Processor and the Context Model Generator. In this demo paper, we presented our implementation of these tools: xPACE and TASC Modeler.

As future work, we plan to apply the tools in different projects and evaluate their quality attributes, in particular time behavior, capacity, and scalability when they have to deal with much bigger contextual datasets.

\bibliographystyle{IEEEtran}
\bibliography{IEEEabrv,_samplepaper}

\end{document}